\documentclass[aps,prl,twocolumn,groupedaddress,showpacs]{revtex4}
\usepackage{graphicx}

\newcommand{\delete}[1]{}

\newcommand{\be}{\begin{equation}}
\newcommand{\ee}{\end{equation}}

\def\beq{\begin{equation}}
\def\eeq{\end{equation}}
\def\bea{\begin{eqnarray}}
\def\eea{\end{eqnarray}}
\def\ba{\begin{array}}
\def\ea{\end{array}}

\begin{document}

\title{Short-range force detection using optically-cooled levitated microspheres}
\author{Andrew A. Geraci}
\email[]{aageraci@boulder.nist.gov}
\author{Scott B. Papp}
\author{John Kitching}
\affiliation{Time and Frequency Division, National Institute of
Standards and Technology, Boulder, CO 80305 USA}

\date{\today}
\begin{abstract}

We propose an experiment using optically trapped and cooled
dielectric microspheres for the detection of short-range forces. The
center-of-mass motion of a microsphere trapped in vacuum can
experience extremely low dissipation and quality factors of
$10^{12}$, leading to yoctonewton force sensitivity.  Trapping the
sphere in an optical field enables positioning at less than $1$
$\mu$m from a surface, a regime where exotic new forces may exist.
We expect that the proposed system could advance the search for
non-Newtonian gravity forces via an enhanced sensitivity of
$10^5-10^7$ over current experiments at the $1$ $\mu$m length scale.
Moreover, our system may be useful for characterizing other
short-range physics such as Casimir forces.

\end{abstract}


\maketitle

Since the pioneering work of Ashkin and coworkers in the 1970s
\cite{ashkin1}, optical trapping of dielectric objects has become an
extraordinarily rich area of research. Optical tweezers are used
extensively in biophysics to study and manipulate the dynamics of
single molecules, and in soft condensed-matter physics to study
macromolecular interactions \cite{grier,block}. Much recent work has
focused on trapping in solution where strong viscous damping
dominates particle motion. There has also been interest in extending
the regime that Ashkin and coworkers opened, namely, trapping
sub-wavelength particles in vacuum where particle motion is strongly
decoupled from a room-temperature environment
\cite{ashkin1,beadexpts}.

Recent theoretical studies have suggested that nanoscale dielectric
objects trapped in ultrahigh vacuum might be cooled to their ground
state of (center-of-mass) motion via radiation pressure forces of an
optical cavity \cite{kimble,cirac}. This remarkable result is made
possible by isolation from the thermal bath, robust decoupling from
internal vibrations, and lack of a clamping mechanism. In fact, a
trapped dielectric nanosphere has been predicted to attain an
ultrahigh mechanical quality factor $Q$ exceeding $10^{12}$ for the
center-of-mass mode, limited by background gas collisions. Such
large $Q$ factors enable cavity cooling, for which the lowest
possible phonon occupation of the mechanical oscillator is $n_T/Q$,
where $n_T$ is the number of room-temperature thermal phonons.
Although such $Q$ factors have yet to be observed in experiment,
optically levitated microspheres have been trapped in vacuum for
lifetimes exceeding $1000$ s \cite{ashkin1} and electrically
levitated microspheres have exhibited pressure-limited damping that
is consistent with theoretical predictions down to $\sim 10^{-6}$
Torr \cite{kendall}.

In addition to being beneficial for ground-state cooling and studies
of quantum coherence in mesoscopic systems, mechanical oscillators
with high quality factors also enable sensitive resonant force
detection \cite{rugar2,yocto}. Optically levitated microspheres in
vacuum have been studied theoretically in the context of reaching
and exceeding the standard quantum limit of position measurement
\cite{libbrecht}. In this paper, we discuss the force sensing
capability of a microsphere trapped inside a medium-finesse optical
cavity at ultra-high vacuum, and propose an experiment that could
extend the search for non-Newtonian gravity-like forces that may
occur at micron scale distances.  Such forces could be mediated by
particles residing in sub-millimeter scale extra spatial dimensions
\cite{add} or by moduli in the case of gauge-mediated supersymmetry
breaking \cite{sg}.  The apparatus we propose is also well suited to
studying Casimir forces \cite{Casimir}, and may be useful for
studying radiative heat transfer at the nano-scale \cite{heatxfer}.

Corrections to Newtonian gravity at short range are generally
parameterized according to a Yukawa-type potential \be
V=-\frac{G_Nm_1m_2}{r}\left[1+\alpha e^{-r/\lambda}\right],
\label{graveq}\ee where $m_1$ and $m_2$ are two masses interacting
at distance $r$, $\alpha$ is the strength of the potential relative
to gravity, and $\lambda$ is the range of the interaction. For two
objects of mass density $\rho$ and linear dimesion $\lambda$ with
separation $r \approx \lambda$, a Yukawa-force scales roughly as
$F_Y \sim G_N \rho^2 \alpha \lambda^4$, decreasing rapidly with
smaller $\lambda$.  For example, taking gold masses, for an
interaction potential with $\alpha=10^5$ and $\lambda=1$ $\mu$m,
$F_Y \sim 10^{-21}$ N. As we will discuss, the thermal-noise-limited
force sensitivity of micron scale, optically levitated silica
micro-spheres at $300$ K with $Q=10^{12}$ can be of order $\sim
10^{-21}$ N$/\sqrt{{\rm{Hz}}}$, and therefore allows probing deep
into unexplored regimes.  For instance, current experimental limits
at $\lambda=1$ $\mu$m have ruled out interactions with $|\alpha|$
exceeding $10^{10}$.

\begin{figure}[!t]
\begin{center}
\includegraphics[width=0.8\columnwidth]{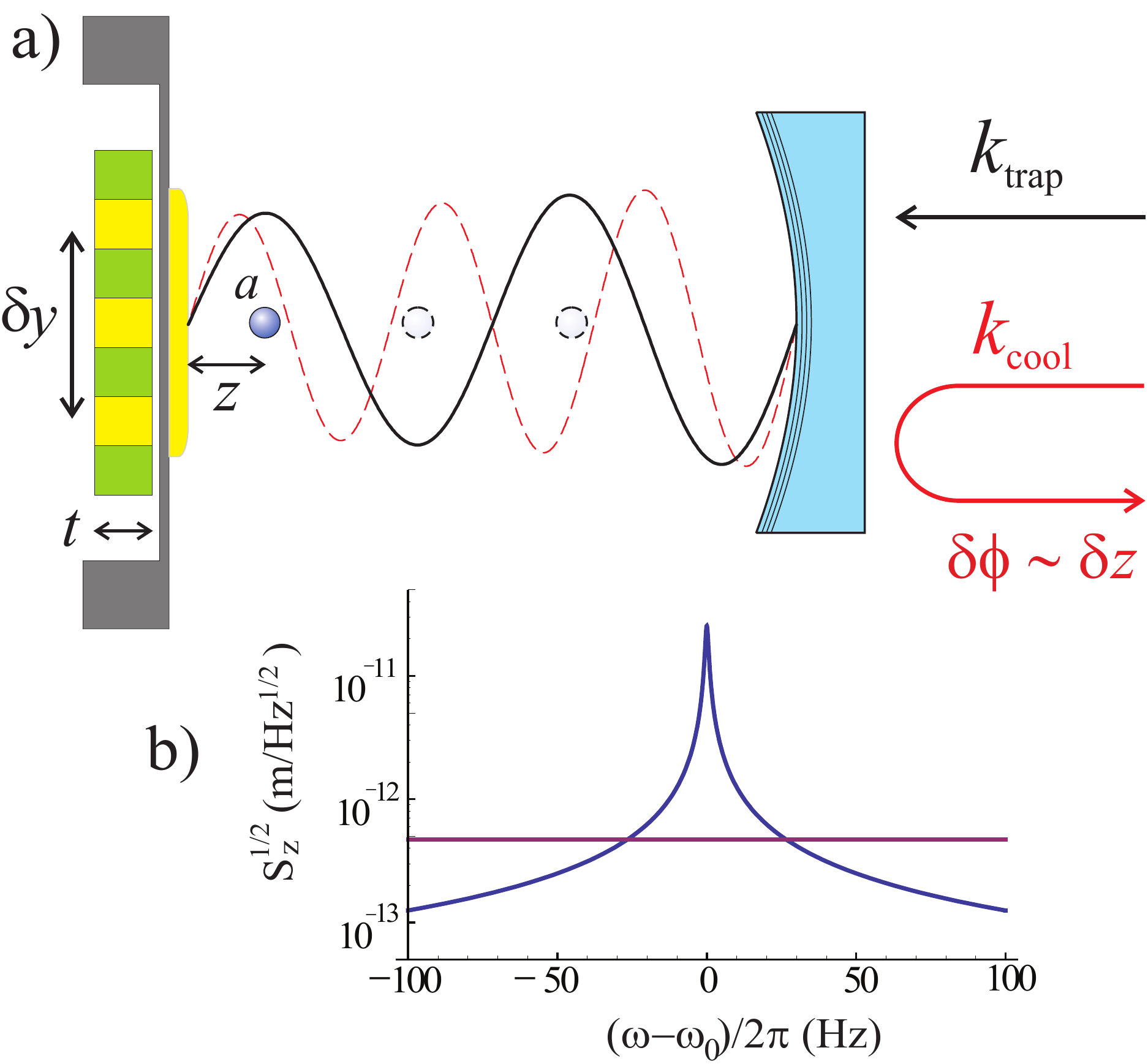}
\caption{(color online) a) Proposed experimental geometry.  A
sub-wavelength dielectric microsphere of radius $a$ is trapped with
light in an optical cavity. The sphere is positioned at an anti-node
occurring at distance $z$ from a gold-coated SiN membrane. Light of
a second wavelength $\lambda_{\rm{cool}}=2 \lambda_{\rm{trap}}/3$ is
used to simultaneously cool and measure the center of mass motion of
the sphere. The sphere displacement $\delta z$ results in a phase
shift $\delta \phi$ in the cooling light reflected from the cavity.
For the short-range gravity measurement, a source mass of thickness
$t$ with varying density sections is positioned on a moveable
element behind the mirror surface that oscillates harmonically with
an amplitude $\delta y$. The source mass is coated with a thin layer
of gold to provide an equipotential. (b) Displacement spectral
density (blue) due to thermal noise and shot-noise limited
displacement sensitivity (flat line, red) for parameters discussed
in the text. \label{cavfig}} \vspace{-20pt}
\end{center}
\end{figure}

The proposed experimental setup is shown schematically in Fig.
\ref{cavfig}.  A dielectric microsphere of radius $a=150$ nm is
optically levitated and cooled in an optical cavity of length $L$ by
use of two light fields of wavenumbers
$k_t=2\pi/\lambda_{\rm{trap}}$ and $k_c=2\pi/\lambda_{\rm{cool}}$,
respectively. The silica microsphere has density $\rho=2300$
kg/m$^3$, dielectric constant $\epsilon=2$, and is trapped near the
position of the closest anti-node of the cavity trapping field to a
gold mirror surface. The mirror is a $200$ nm thick SiN membrane
coated with $200$ nm of gold.  A source mass of thickness $t=5$
$\mu$m and length $20$ $\mu$m with varying density sections of width
$2$ $\mu$m (e.g., Au and Si) is positioned at edge-to-edge
separation $d=1$ $\mu$m from the sphere.  Below we describe trapping
and cooling of the microsphere's center-of-mass motion, detection of
Casimir forces between the microsphere and gold mirror, and the
search for gravity-like forces on the microsphere due to the source
mass.

Following Ref. \cite{kimble}, the sub-wavelength dielectric particle
has a center-of-mass resonance frequency $ \omega_0 = \left[
\frac{6k_t^2I_t}{\rho c} {\mathcal{R}}e
\frac{\epsilon-1}{\epsilon+2} \right]^{1/2}$, where $I_t$ is the
intracavity intensity of the trapping light. The trap depth is
$U=\frac{3I_{t}V}{c} \frac{\epsilon-1}{\epsilon+2}$, where $V$ is
the volume of the microsphere.  For concreteness, we consider a
cavity of length $L=0.15$ m, finesse ${\mathcal{F}}=200$, driven
with a trapping laser of power $P_t=2$ mW and wavelength
$\lambda_{\rm{trap}}=1.5$ $\mu$m. We choose a cavity mode waist
$w=15$ $\mu$m.   The Gaussian profile of the trapping beam near the
mode waist provides transverse confinement, with an oscillation
frequency of $\sim 1$ kHz.  Tighter transverse confinement could be
established by use of a transverse standing wave potential. The
cooling light has input power $P_c=48$ $\mu$W, and an optimized red
detuning of $\delta =-0.23 \kappa$, where the cavity decay rate is
$\kappa = \pi c / L {\mathcal{F}}$. The cooling light causes a
slight shift $z_0$ in the axial equilibrium position of the
microsphere, given by $ z_0=\frac{1}{2}\frac{k_c}{k_t^2}
\frac{I_c}{I_t} \approx 2$ nm, where $I_c$ is the intracavity
intensity of the cooling mode. The optomechanical coupling of the
cooling mode is $ g=\frac{3V}{4V_c} \frac{\epsilon-1}{\epsilon+2}
\omega_c,$ where $V_c=\pi w^2 L /4$ is the cavity mode volume
\cite{kimble}, and $\omega_c=k_c c$. The optimum detuning is
determined by minimizing the final phonon occupancy $n_f$, which
depends on the laser-cooling rate and heating due to photon recoil
from light scattered by the sphere. Additional cavity loss due to
photon scattering is negligible: $\sim 10^{-3} \kappa$ for our
parameters. Values of the trapping and cooling parameters appear in
Table \ref{table1}.

A mechanical oscillator with frequency $\sim 37$ kHz and $Q \sim
10^{12}$ will respond to perturbations with a characteristic time
scale of $2Q/\omega_0 \sim 10^7$ s. The cooling serves both to damp
the $Q$ factor so that perturbations to the system ring down within
reasonably short periods of time, and to localize the sphere by
reducing the amplitude of the thermal motion. Because of the low
cavity finesse, the cooling is not in the resolved sideband regime.
Still, for the parameters discussed above the phonon occupation of
the microsphere oscillation is reduced by a factor of over $10^5$.
This corresponds to operating with an effective $Q_{\rm{eff}}
\approx 10^5$ and ring down time of $\approx 1$ s. Cooling of the
transverse motion is also required, as the rms position spread must
be maintained to be less than $\sim 0.1$ $\mu$m. We imagine this can
be done with active feedback to modulate the power of a transverse
trapping laser using the signal from a transverse position
measurement, for example generated by measuring scattered light
incident on a quadrant photodiode.  A modest cooling factor of
$\approx 1000$ in the transverse directions is sufficient to yield
the required localization.

The cooling light is also used to detect the position of the sphere.
The phase of the cooling light reflected from the cavity is
modulated by microsphere motion through the optomechanical coupling
$\partial{\omega_c}/\partial{z}=2 k_c g$. Photon shot-noise limits
the minimum detectable phase shift to $\delta \phi \approx
1/(2\sqrt{I})$ where $I \equiv P_c /(\hbar \omega_c)$ \cite{hadjar}.
The corresponding photon shot-noise limited displacement sensitivity
 is $ \sqrt{S_z(\omega)}= \frac{\kappa}{4k_cg}
\frac{1}{\sqrt{I}}\sqrt{1+\frac{4\omega^2}{\kappa^2}}$
\cite{hadjar}, for an impedance matched cavity.  This displacement
sensitivity is generally well below the thermal noise limited
sensitivity, as shown in Fig. \ref{cavfig}(b). We assume that
substrate vibrational noise, electronics noise and laser noise can
be controlled at a level comparable to the photon shot noise. The
minimum detectable force due to thermal noise at temperature
$T_{\rm{eff}}$ is $ F_{\rm{min}} = \sqrt{\frac{4 k k_B T_{\rm{eff}}
b}{\omega_0 Q_{\rm{eff}}}}$, where $k$ is the center-of-mass mode
spring constant, and $b$ is the bandwidth of the measurement. We
assume an initial center-of-mass temperature $T_{\rm{CM}}=300$ K,
and that $Q \approx \omega_0 / \gamma_g$ is limited by background
gas collisions, with loss rate $\gamma_g=16 P_{\rm{gas}}/(\pi
\bar{v} \rho a)$ \cite{epstein}, for a background air pressure of
$P_{\rm{gas}}=10^{-10}$ Torr and rms gas velocity $\bar{v}$. Cooling
the center-of-mass mode to $T_{\rm{eff}}= 0.9$ mK results in
$F_{\rm{min}} \sim 10^{-21}$ N$/\sqrt{\rm{Hz}}$ as shown in Table
\ref{table1}. In this regime $F_{\rm{min}}$ scales linearly with the
sphere radius $a$.

The microsphere absorbs optical power from both the trapping and
cooling light in the cavity, which results in an increased internal
temperature $T_{\rm{int}}$.  Assuming negligible cooling due to gas
collisions, the absorbed power is re-radiated as blackbody
radiation.  $T_{\rm{int}}$ is listed in Table \ref{table1} for fused
silica with dielectric response $\epsilon=\epsilon_1+i\epsilon_2$,
with $\epsilon_1=2$ and $\epsilon_2=1.0 \times 10^{-5}$ as in Ref.
\cite{fusedsilicaloss}, and $\epsilon_{\rm{bb}}=0.1$ as in Ref.
\cite{kimble}, for an environmental temperature $T_{\rm{ext}}=300$
K. We assume $T_{\rm{int}}$ and $T_{\rm{CM}}$ are not significantly
coupled over the time scale of the experimental measurements at
$P_{\rm{gas}} \sim 10^{-10}$ Torr.

\begin{table}[!t]
\begin{center}
  \begin{tabular}{@{}ccc@{}}
  \hline
  \hline
  Parameter & Units & Value \\
  \hline
$\lambda_{\rm{trap}}$ & $\mu$m & $1.5$ \\
$U/k_B$ & K & $3.7 \times 10^3 $ \\
$\omega_0/2\pi$ & Hz & $3.7 \times 10^4$  \\
$ T_{\rm{int}} $ & K & $900$ \\
$ Q,(Q_{\rm{eff}}) $ & - & $6.1 \times 10^{11},(1.0 \times 10^{5})$ \\
$\delta/\kappa$ & - & $-0.23$ \\
$ n_T,(n_f) $ & - & $1.7 \times 10^8$,$(510)$ \\
$ \sqrt{S_z} $ & m$/\sqrt{\rm{Hz}}$ & $4.7 \times 10^{-13}$ \\
$F_{\rm{min}}$ & N$/\sqrt{\rm{Hz}}$ & $1.9 \times 10^{-21} $ \\
$ z_{\rm{th}} $ & m$/\sqrt{\rm{Hz}}$ & $2.6 \times 10^{-11} $ \\
  \hline
  \hline
  \end{tabular}
\caption{\label{table1} Parameters for trapping and cooling a silica
sphere with radius $a=150$ nm.} \vspace{-20pt}
\end{center}
\end{table}

{\it{Casimir Force.}}  The Casimir force between a dielectric sphere
and metal plane can be written using the proximity force
approximation (PFA) as \cite{Casimir} $F_{\rm{c}}=- \eta \frac{\pi^3
a \hbar c} {360 (z-a)^3}$ in the limit that $(z-a) \ll a$.   The
prefactor $\eta$ characterizes the reduction in the force compared
with that between two perfect conductors \cite{lambrecht}.  For $z
\gg a$, the force takes the Casimir-Polder \cite{casimirpolder} form
$F_{\rm{cp}}=-\frac{3\hbar c \alpha_V}{8\pi^2\epsilon_0}
\frac{1}{z^5}$, where $\alpha_V=3\epsilon_0 V
\frac{\epsilon-1}{\epsilon+2}$ is the electric polarizability.  Our
setup is capable of probing the unexplored transition between these
two regimes, and of testing the PFA, which is expected to be valid
for $z-a \lesssim a$ \cite{jaffe}. To estimate $\eta$, we adopt a
similar approach to that taken in Ref. \cite{lambrecht} to determine
the force between a metal and dielectric plate. We assume dielectric
spheres separated from a metal mirror will have a similar
pre-factor.  Taking an infinite plate with $\epsilon=2$ and
thickness $2a$, and another with gold of thickness $200$ nm, we find
$\eta \approx 0.13$ at $(z-a)=225$ nm.

\begin{figure}[!t]
\begin{center}
\includegraphics[width=1.0\columnwidth]{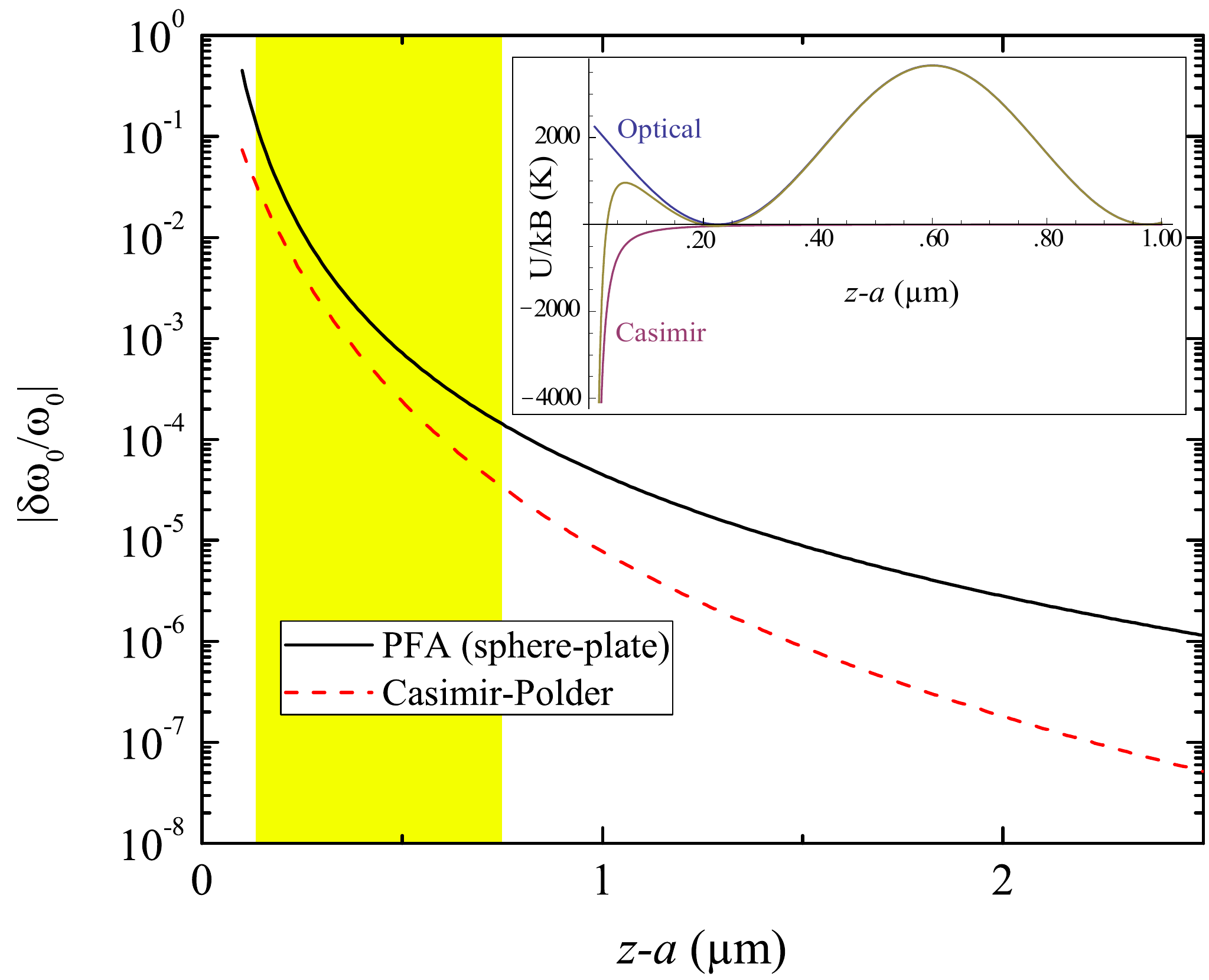}
\caption{(color online) Fractional frequency shift due to Casimir
interaction of microspheres of radius $a=150$ nm at distance $z$
from the gold surface.  The PFA is expected to be valid at short
distances, while the Casimir-Polder form is expected at large
distances, and the transition region is shown as a shaded area.
(inset) Optical and Casimir contributions to the cavity trapping
potential. \label{freqshift}} \vspace{-20pt}
\end{center}
\end{figure}

For a sphere located at the position of the first anti-node of the
trapping field, the Casimir force displaces the equilibrium position
by approximately $-3$ nm. The gradient of the Casimir force near the
static mirror surface produces a fractional shift in the resonance
frequency of the sphere given by $|\delta \omega_0 / \omega_0| =
\frac{|\partial{F_{\rm{c}}}/\partial{z}|}{2k}$. A similar frequency
shift has been measured for an atomic sample \cite{harber}. The
shift is shown in Fig. \ref{freqshift} as a function of mirror
separation $(z-a)$ for $\eta=0.13$.  The minimum detectable
frequency shift due to thermal noise is given by $|\delta \omega_0 /
\omega_0|_{\rm{min}}=\sqrt{\frac{k_BT^{\rm{f}}_{\rm{CM}}
b}{k\omega_0Q_{\rm{eff}}z_{\rm{rms}}^2}}$. For $z_{\rm{rms}}=5$ nm,
$|\delta \omega_0 / \omega_0| \approx 10^{-7}$ can be detected in
$\sim 1$ s. Other sources of systematic frequency shifts near the
surface, for example from variation of the cavity finesse with bead
position or from diffuse scattered light on the gold surface, would
need to be experimentally characterized. Also, surface roughness of
the microsphere can modify the Casimir force \cite{neto}. Rotation
of the microsphere may lead to an effective averaging over surface
inhomogeneities.

\begin{figure}[!t]
\begin{center}
\includegraphics[width=1.0\columnwidth]{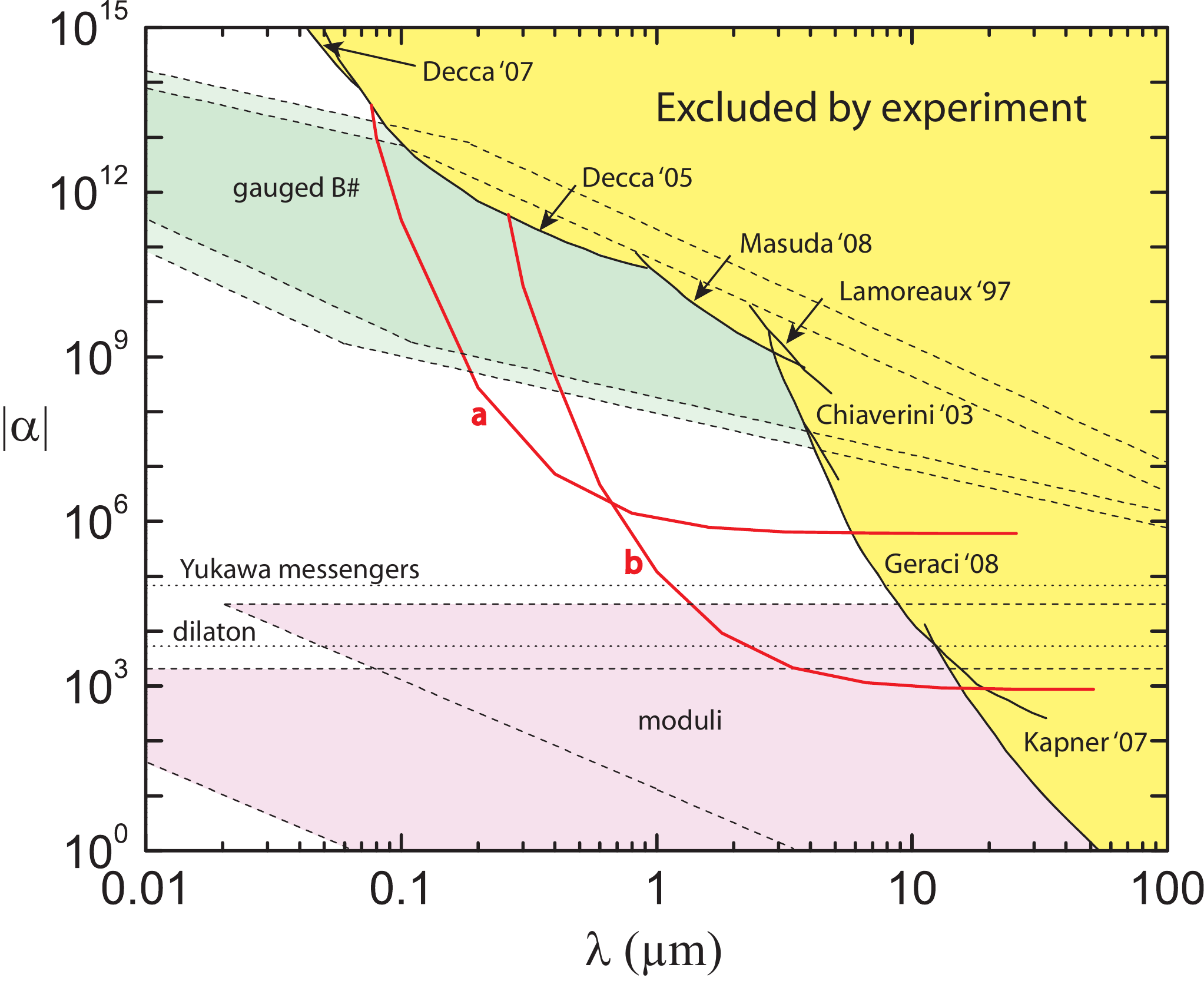}
\caption{(color online) Experimental constraints
\cite{decca07,decca05,masuda08,lamoreaux97,chiaverini03,geraci08,kapner07}
and theoretical predictions \cite{bec} for short-range forces due to
an interaction potential of Yukawa form
$V=-\frac{G_Nm_1m_2}{r}\left[1+\alpha e^{-r/\lambda}\right]$. Lines
(a) and (b) denote the projected improved search reach for
microspheres of radius $a=150$ nm and $a=1500$ nm, respectively.
\label{alam}}\vspace{-20pt}
\end{center}
\end{figure}

{\it{Search for non-Newtonian gravity.}} To generate a modulation of
any Yukawa-type force at the resonance frequency of the
center-of-mass mode along $z$, the source mass is mounted on a
cantilever beam that undergoes a lateral tip displacement of 2.6
$\mu$m at a frequency of $\omega_0 /3$.  The mechanical motion
occurs at a sub-harmonic of the microsphere resonance to avoid
direct vibrational coupling. To estimate the force on the sphere a
numerical integration over the geometry of the masses is performed.
For $b=10^{-5}$ Hz, the estimated search reach is shown in Fig.
\ref{alam} (a). Several orders of magnitude of improvement are
possible between $0.1$ nm and a few microns, due to the proximity of
the masses and high force sensitivity.

The source mass surface is coated with $200$ nm of gold in order to
screen the differential Casimir force, depending on which material
is directly beneath the microsphere.  Following the method of Ref.
\cite{lambrecht}, the differential Casimir force is $9 \times
10^{-24}$ N, which is comparable to the sensitivity of the
experiment at $10^{-5}$ Hz bandwidth. The gold coating on the cavity
mirror membrane attenuates this Casimir interaction even further.
Patch potentials on the mirror surface and any electric charge on
the sphere can produce spurious forces on the sphere.  By
translating the position of the optical trap along the surface,
these and other backgrounds, e.g. vibration, can be distinguished
from a Yukawa-type signal, as any Yukawa-type signal should exhibit
a spatial periodicity associated with the alternating density
pattern, similar to the system discussed in Ref. \cite{geraci08}.

Increasing the radius of the sphere can significantly enhance the
search for non-Newtonian effects at longer range. Curve (b) in Fig.
\ref{alam} shows the estimated search reach that would be obtained
by scaling the sphere size by a factor of 10 and positioning it at
edge-to-edge separation of $3.8$ $\mu$m from a source mass with
thickness $t=10$ $\mu$m consisting of sections of width $10$ $\mu$m
driven at an amplitude of $13$ $\mu$m. Such a larger sphere could be
trapped in an optical lattice potential with the incident beams at a
shallow angle, instead of in an optical cavity, to enable
sub-wavelength confinement. In this case cooling could be performed
by use of active feedback. Alternatively it may be possible to trap
the larger $1.5$ $\mu$m sphere in a cavity by use of longer
wavelength light (e.g., $\lambda_{\rm{trap}} = 10.6$ $\mu$m) by
choosing a sphere material such as ZnSe with lower optical loss at
this wavelength.

The experiment we have proposed may allow improvement by several
orders of magnitude in the search for non-Newtonian gravity below
the $10$ $\mu$m length scale.  An experimental challenge will be to
capture and cool individual microspheres and precisely control their
position near a surface.  Previous experimental work has been
successful at optically trapping $1.5$ $\mu$m radius spheres
\cite{beadexpts}, and similar techniques may work for the setup
proposed here. Extrapolating the results of Ref. \cite{kendall} at
$10^{-6}$ Torr for the system we consider would yield a
pressure-limited $Q \sim 10^9$. In the absence of additional damping
mechanisms, we expect that $Q \approx 10^{12}$ could be achieved at
lower pressure. Further improvements in force sensitivity may be
possible in a cryogenic environment.

We thank John Bollinger and Jeff Sherman for a careful reading of
this manuscript. AG and SP acknowledge support from the NRC.

\end{document}